\documentclass[twocolumn]{autart}
\usepackage{amsmath}
\usepackage{amssymb}
\usepackage{xcolor}
\usepackage{graphicx}
\usepackage[round]{natbib}
\usepackage{enumerate}
\renewcommand{\qed}{\hfill \mbox{$\blacksquare$}}

\def \R{{\mathbb R}}
\def \N{{\mathbb N}}

\def\K{\mathcal{K}}
\def\Ki{\mathcal{K}_{\infty}}
\def\C{\mathcal{C}}
\def\KL{\mathcal{KL}}
\def\U{\mathcal{U}}

\begin{document}

\begin{frontmatter}

\title{%Forward completeness is not robust for time-delay systems\thanksref{footnoteinfo}}
%Time-delay systems that defy intuition:
%nonrobust forward completeness and related (non)properties
%
Forward completeness does not imply bounded reachability sets and global asymptotic stability is not necessarily uniform for time-delay systems
\thanksref{footnoteinfo}}
%
% and uniform global attractivity and Lyapunov stability do not imply uniform global asymptotic stability

% the curse of infinite dimensionality cannot be overcome
% 
% Local Stability (LS)
% FC y RFC

\thanks[footnoteinfo]{Partially supported by Agencia I+D+i grants PICT 2018-1385, 2021-0730, Argentina.}

\author[JLaddress]{Jose L. Mancilla-Aguilar}\ead{jmancil@fi.uba.ar},
\author[HHaddress]{Hernan Haimovich}\ead{haimovich@cifasis-conicet.gov.ar}

\address[JLaddress]{Departamento de Matem\'atica, Facultad de Ingenier\'{\i}a, Universidad de Buenos Aires, Av. Paseo Colon 850, CABA, Argentina.}
\address[HHaddress]{International French-Argentine Center for Information and Systems Science (CIFASIS),\\ CONICET-UNR, Ocampo y Esmeralda, 2000 Rosario, Argentina.}


\begin{abstract}
  An example of a time-invariant time-delay system that is uniformly globally attractive and exponentially stable, hence forward complete, but whose reachability sets from bounded initial conditions are not bounded over compact time intervals is provided. This gives a \emph{negative answer} to two current conjectures by showing that (i) forward completeness is not equivalent to robust forward completeness (i.e. boundedness of reachability sets) and (ii) global asymptotic stability is not equivalent to uniform global asymptotic stability. In addition, a novel characterization of robust forward completeness for %usually encountered classes of time-delay systems
  systems having a finite number of discrete delays is provided. This characterization relates robust forward completeness of the time-delay system with the forward completeness of an associated nondelayed finite-dimensional system.
\end{abstract}

\end{frontmatter}
%\maketitle

\section{Introduction}
\label{sec:introduction}

Forward completeness is the name given to the property of a dynamical system whose solutions exist for all future times. Standard solution concepts entail the fact that a solution remains bounded on every compact time interval contained in its maximum interval of existence. The fact that every solution is bounded on compact time intervals does not a priori imply that uniform bounds will exist for solutions beginning from bounded sets of initial conditions and/or corresponding to uniformly bounded inputs. For time-invariant finite-dimensional systems, \citet{linson_jco96} showed that under reasonable (Lipschitz) continuity assumptions such uniform bounds exist and hence forward completeness becomes equivalent to the boundedness of the reachability sets of solutions beginning from bounded sets of initial conditions and corresponding to uniformly bounded inputs, over compact time intervals.

For infinite-dimensional systems, the situation can be radically different. \citet[Examples~2 and~3]{mirwir_tac18} showed that forward completeness (even in addition to other properties) does not imply the aforementioned boundedness of the reachability sets. The cause for this difference between what happens for finite- and infinite-dimensional systems can be traced back to the fact that compactness being equivalent to closedness plus boundedness is used in an essential way in the proof given by \citet[Proposition~5.1]{linson_jco96}, and this equivalence ceases to hold for infinite-dimensional systems. The state space of Examples~2 and~3 of \citet{mirwir_tac18} is the set $l_2$ of square-summable sequences and has no obvious relationship to time-delay systems.

In the context of time-delay systems, the aforementioned boundedness of reachability sets property may be called robust forward completeness %% \footnote{The definition of robust forward completeness in \citet[Definition~2.1]{karjia_book11} is different and stronger, but coincides with that of \citet{chakar_mcss23} in the absence of inputs.}
\citep{karjia_book11,chakar_mcss23}.
Although being essentially infinite-dimensional, time-delay systems may have their own unique features which do not necessarily hold for other types of infinite-dimensional systems \citep{halver_book93}. Therefore, the question of whether forward completeness could be equivalent to robust forward completeness remained open \citep[Conjecture~1]{chakar_mcss23}, until now. 

For time-invariant finite-dimensional input-free systems, it is also known that global asymptotic stability, understood as the combination of local Lyapunov stability and every trajectory converging to zero, is equivalent to uniform global asymptotic stability, where the norm of the state trajectory is bounded by some $\KL$-function of the initial state and elapsed time \citep[see page 109 of][Theorems~21 and 23]{hahn_book67,masser_am56}. For time-delay systems, by contrast, the validity of this equivalence also remained unknown \citep[Conjecture~2]{chakar_mcss23}. 
%% It was very recently proved that GAS jointly with RFC is equivalent to UGAS for time-invariant time-delay input-free systems \citep[Theorem~1]{karpep_jco22}. The question of whether GAS by itself is equivalent to UGAS also remained open \citep{karpep_jco22, chakar_mcss23}.

%% Add GAS and UGAS

In this note, a forward-complete, exponentially stable (hence Lyapunov stable) and uniformly globally attractive time-delay system that does not have the boundedness of reachability sets property is constructed.
The constructed system has a single discrete delay, no external inputs, and in addition has the highly counterintuitive properties of being uniformly globally attractive and globally asymptotically stable, but not uniformly globally asymptotically stable. This gives a definite negative answer to the above two conjectures.
In order to prove the specific properties and lack of properties of the constructed system, novel characterizations of (robust) forward completeness are provided as a side contribution.

%% According to \citet[Theorem~1]{karpep_jco22} and since the constructed system is not robustly forward complete, then the system cannot be uniformly globally asymptotically stable, meaning that no uniform decaying bound over all trajectories starting from bounded initial conditions exists. This also gives a negative answer to Conjecture~2 of \citet{chakar_mcss23}, where the equivalence between global asymptotic stability and its uniform version was conjectured for time-delay systems.

\section{Basic Definitions and Assumptions}
\label{sec:basic-definitions}

\subsection{Notation}
\label{sec:notation}
$\N$ denotes the set of positive integers, $\R$ denotes the reals and $\R_{\ge 0}=[0,\infty)$. % and $\R_{>0}=(0,\infty)$.
  We write $\alpha\in\K$ if $\alpha:\R_{\ge 0} \to \R_{\ge 0}$ is continuous, strictly increasing and $\alpha(0)=0$, and $\alpha\in\Ki$ if, in addition, $\alpha$ is unbounded. We write $\beta\in\KL$ if $\beta:\R_{\ge 0}\times \R_{\ge 0}\to \R_{\ge 0}$, $\beta(\cdot,t)\in \K$ for any $t\ge 0$ and, for any fixed $r\ge 0$, $\beta(r,t)$ monotonically decreases to zero as $t\to \infty$.
  For $k\in \N$ and $y\in\R^k$, $|y|=|y|_{\infty}=\max\{|y_i|:i=1,\ldots,k\}$ and $|y|_2=\sqrt{y^Ty}$ denote, respectively, the supremum and Euclidean norms of $y$. Given $r\ge 0$, $B_r^k = \{ y \in \R^k : |y| \le r \}$. 
We denote by $\U^r$ the set of locally essentially bounded Lebesgue measurable functions $u:\R_{\ge 0}\to \R^r$, whose elements will be referred to as inputs. We identify $\U_c^r \subset \U^r$ as the subset of continuous inputs. For an input $u$, $\|u\|_{\infty}={\rm ess\;sup}_{t\ge 0}|u(t)|$ and given an interval $I\subset \R_{\ge 0}$, $u_I$ denotes the input that satisfies $u_{I}(t)=u(t)$ for all $t\in I$ and $u_{I}(t)=0$ elsewhere. Note that $\|u_{I}\|_{\infty}<\infty$ for every bounded interval $I$.
Given $\tau\ge 0$,
%(where $\tau$ is larger than, or equal to, the maximum delay involved in the dynamics),
let $\mathcal{C}=\mathcal{C}\left([-\tau,0],\mathbb{R}^{n}\right)$ be the set of continuous functions $\psi:[-\tau,0]\to \R^n$ endowed with the supremum norm $\|\psi\|=\displaystyle\max\{|\psi(s)|:s\in [-\tau ,0]\}$. As usual, given a continuous function $x:[t_0-\tau,T)\to \R^n$ and any $t_0\le t<T$, $x_t$ is defined as the function $x_t:[-\tau,0]\to \R^n$ satisfying $x_{t}(s)=x(t+s)$ for all $s\in [-\tau,0]$, so that $x_t\in \mathcal{C}$. For $r\ge 0$, we define $B_r^{\C} := \{ \psi \in \mathcal{C} : \|\psi\| \le r \}$.
%% A continuous function $V:\R^n\to \R_{\ge 0}$ is proper if every sublevel set $V^{-1}([0,c])=\{\xi\in \R^n:V(\xi)\le c\}$ is compact or, equivalently, $\lim_{|\xi|\to \infty}V(\xi)=\infty$; $V$ is smooth if $V\in \mathcal{C}^{\infty}(\R^n)$; and the gradient of $V$ at $\xi \in \R^n$ is denoted by $\nabla V(\xi)$.
  We say that a function $f:\C\times \R^m\to \R^n$ is
  %completely continuous if it is continuous and bounded on bounded subsets of $\C\times \R^m$ and that it is
  Lipschitz on bounded sets if for each $R\ge 0$ there exists $L\ge 0$ such that for all $\phi,\psi\in B_R^{\C}$ and $\mu, \nu \in B_R^m$, it happens that
\begin{align}
|f(\phi,\mu)-f(\psi, \nu)|\le L (\|\phi-\psi\|+|\mu-\nu|).
\end{align}  

\subsection{Time-delay Systems}
\label{sec:sys-cons}
  
Consider the time-invariant system with inputs defined by the following retarded functional differential equation
\begin{align}\label{eq:sistr}
\dot{x}(t)=f(x_{t},u(t))
\end{align}
where $t\ge 0$, $x(t)\in \R^n$, $u(t) \in \R^m$, $x_t \in \C$ and $\mathcal{C}$ as defined as in Section~\ref{sec:notation} %$x_{t}(s)=x(t+s)$ for all $s\in [-\tau,0]$, 
with $\tau\ge 0$ not less than the maximum delay involved in the dynamics. %and $\mathcal{C} = \mathcal{C}\left([-\tau,0],\mathbb{R}^{n}\right)$.
\begin{assum}\label{as:Lipf}
  The function $f:\mathcal{C}\times\R^{m}\to\R^{n}$ in~(\ref{eq:sistr}) is Lipschitz on bounded sets.
\end{assum}

For an input $u \in \U^m$, a solution of (\ref{eq:sistr}) is a continuous function $x:[-\tau,t_f)\to \R^n$, with $t_f>0$ or $t_f = \infty$, that is locally absolutely continuous on $[0,t_f)$ and such that $\dot{x}(t)=f(x_t,u(t))$ for almost all $t\in [0,t_f)$, i.e.
\begin{align}\label{eq:int}
x(t)=x(0)+\int_{0}^t f(x_s,u(s))ds \quad \forall t\in [0,t_f).
\end{align}
%% Since any solution $x:[-\tau,t_f)\to \R^n$ is a continuous function of time, then $x$ is bounded over any compact time interval contained within $[-\tau,t_f)$, i.e. 
%%     \begin{align*}
%%       -\tau \le a < b < t_f \quad\Rightarrow\quad \sup_{t\in [a,b]} |x(t)| < \infty.
%%     \end{align*}
 A solution $x$ is maximally defined if there does not exist another solution $\tilde x:[-\tau,\tilde t_f)$ of (\ref{eq:sistr}) such that $\tilde t_f>t_f$ and $x(t)=\tilde x(t)$ for all $t\in [-\tau,t_f)$.

Under Assumption~\ref{as:Lipf}, for every $\phi \in \mathcal{C}$ and $u \in \U^m$, there exists a unique maximally defined solution $x:[-\tau,t_{(\phi,u)})\to \R^n$ of (\ref{eq:sistr}) that satisfies $x_0=\phi$; in addition, system (\ref{eq:sistr}) exhibits the boundedness implies continuation (BIC) property, {\em i.e.} $t_{(\phi,u)}<\infty$ implies that $x(\cdot)$ is unbounded \citep[see e.g.][]{halver_book93}. We denote by $x(\cdot,\phi,u)$ such a unique maximally defined solution. % and by $I_{\phi,u}=[-\tau,t_{(\phi,u)})$ its interval of definition. 
%
%For system (\ref{eq:sistr}) we consider the following properties.
\begin{defn}
  \label{def:fc}  \label{def:fc-cont}  \label{def:rfc}
  System (\ref{eq:sistr}) is 
  \begin{itemize}
  \item {\em forward complete} (FC) if $t_{(\phi,u)}=\infty$ for all $\phi \in \C$ and $u\in \U^m$;
  \item {\em FC for continuous inputs} (FCC) if $t_{(\phi,u)}=\infty$ for all $\phi \in \C$ and $u\in \U^m_c$;
  \item {\em robustly forward complete} (RFC) if it is FC and for every $R\ge 0$ and $T\ge 0$
    \begin{align*} %\label{eq:rfc}
      \sup\{|x(t,\phi,u)|:{\scriptstyle \phi\in B_R^{\C},\;u\in\U^m,\|u\|_{\infty}\le R,\;t\in [0,T]}\}<\infty.
    \end{align*}
  \end{itemize}
\end{defn}
Due to causality, the condition $\|u\|_{\infty}\le R$ can be replaced by $\|u_{[0,T]}\|_{\infty}\le R$. 
In words, system (\ref{eq:sistr}) is FC if its solutions are defined for all future times, for every continuous initial condition and every Lebesgue measurable and locally essentially bounded input. FCC is thus a weaker property because it requires existence of solutions for all future times only for continuous inputs. And clearly, RFC is the strongest of the above three.
%% \begin{defn}
%%   \label{def:fc-cont}
%%   System (\ref{eq:sistr}) is {\em FC for continuous inputs} if $t_{(\phi,u)}=\infty$ for all $\phi \in \C$ and $u\in \U^m_c$.
%% \end{defn}
%% In virtue of the BIC property, this is equivalent to that for every $\phi\in \C$, $u \in \U$ and $T>0$ the following holds
%% \begin{align} \label{eq:fc}
%% \sup\{|x(t,\phi,u)|:\;t\in [0,T] \cap [0,\inftyt_{(\phi,u)})\}<\infty.
%% \end{align}
%% A property that is stronger than FC is the following.
%% \begin{defn} \label{def:rfc} System (\ref{eq:sistr}) is {\em robustly forward complete} (RFC) if it is FC and for every $R\ge 0$ and $T\ge 0$
%% \begin{align} \label{eq:rfc}
%% \sup\{|x(t,\phi,u)|:\phi\in B_R^{\C},\;\|u\|_{\infty}\le R,\;t\in [0,T]\}<\infty.
%% \end{align}
%% \end{defn}
This definition of RFC coincides with that of \citet{chakar_mcss23}, with RFC from the input $u$ in \citet[Definition 1.4]{karjia_book11}, and with the boundedness of reachability sets (BRS) property in \citet{mirwir_tac18}. The RFC property imposes a uniform bound to the solutions of (\ref{eq:sistr}) on compact time intervals provided the initial conditions belong to a bounded set and the inputs are essentially uniformly bounded. %% Note that the definition of RFC in \citet[Definition 2.1]{karjia_book11} is slightly different in the sense that it requires solultions to have a uniform bound over all admissible inputs, not just for inputs in bounded balls, and that the definitions coincide for systems without inputs.
%Therefore, the current definition of RFC is weaker than that of \citet[Definition 2.1]{karjia_book11}.
%The RFC property has important consequences for time-delay systems \citep{chakar_mcss23}.

Next, consider the time-invariant system without inputs
\begin{align}\label{eq:sistrwi}
\dot{x}(t)=f_0(x_{t}).
\end{align}
\begin{assum}
  \label{as:f0Lip}
  The function $f_0 : \C \to \R^n$ is Lipschitz on bounded sets and satisfies $f(\mathbf{0})=0$, with $\mathbf{0}\in \C$ the zero function.
\end{assum}
The unique maximally defined solution of~(\ref{eq:sistrwi}) corresponding to an initial condition $\phi\in\C$ is denoted by $x(\cdot,\phi)$ and its maximal interval of definition by $I_{\phi}=[0,t_{\phi})$. We consider the following stablity properties of the zero solution.
\begin{defn} The zero solution of system (\ref{eq:sistrwi}) is:
\begin{itemize}
\item Lyapunov stable (LS) if for all $\varepsilon>0$ there exists $\delta=\delta(\varepsilon)>0$ such that $|x(t,\phi)|\le \varepsilon$ for all $t\ge 0$, $\phi\in B_{\delta}^{\C}$.
\item Exponentially stable (ES) if there exist $r,p,k>0$ such that $|x(t,\phi)| \le k\|\phi\|e^{-pt}$ for all $t\ge 0$, $\phi\in B_r^{\C}$.
\item Globally attractive (GA)
  if $\lim_{t\to \infty}x(t,\phi)=0$ for all $\phi \in \C$.
\item Uniformly globally attractive (UGA)
  if for every $r>0$ and $\varepsilon > 0$ there exists $T\ge 0$ such that $|x(t,\phi)| \le \varepsilon$ for all $t\ge T$ and $\phi\in B_r^{\C}$.
\item Globally asymptotically stable (GAS)
  if LS and GA.
\item Uniformly globally asymptotically stable (UGAS)
  if there exists $\beta\in \KL$ such that for all $t\ge 0$ and $\phi \in \C$, then
$|x(t,\phi)|\le \beta(\|\phi\|,t)$.
\end{itemize}
System~(\ref{eq:sistrwi}) is said to be LS, ES, GA, UGA, GAS or UGAS if its zero solution has the corresponding property.
\end{defn}
%Note that UGA implies GA, but UGA and GAS together do not imply UGAS (see Corollary~\ref{cor:uga+gas-ugas} in Section~\ref{sec:example-rationale}). 
%% It is known that FC is equivalent to RFC for time-invariant finite-dimensional systems with inputs \citep{linson_jco96} and that the equivalence does not hold for general infinite-dimensional systems \citep{mirwir_tac18}. Whether the equivalence holds for the specific case of time-delay systems remained as an open question \citep{chakar_mcss23}. 
%
%% It is also known that for time-invariant finite-dimensional input-free systems, GAS is equivalent to UGAS \citep{sonwan_tac96}. It was very recently proved that GAS jointly with RFC is equivalent to UGAS for time-invariant time-delay input-free systems \citep[Theorem~1]{karpep_jco22}. The question of whether GAS by itself is equivalent to UGAS also remained open \citep{karpep_jco22, chakar_mcss23}.
%
Note that ES implies LS, and UGA implies GA. 
Next, it will be shown that FC, even in combination with ES and UGA, does not imply RFC and that the combination of UGA and GAS does not imply UGAS.
%This lack of robustness seems to be highly counterintuitive.

\section{Nonrobust Forward Completeness}
\label{sec:FCneqRFC}

\subsection{Example, Main Results and Rationale}
\label{sec:example-rationale}

Consider the planar nondelayed system with a scalar input, defined by
\begin{align}
  \label{eq:planar}
  \dot{x}(t)=\left(1+|x(t)|_2^2\right)A(\varphi(u(t)))x(t)=: g(x(t),u(t))
\end{align}
where $t\ge 0$, $x(t)\in \R^2$, $u(t)\in \R$, $A(\lambda)=\lambda A_1+(1-\lambda)A_2$ for all $\lambda \in \R$,
\begin{align*}\scriptstyle
  \varphi(r) &\scriptstyle = 
  \begin{cases}
    0 \quad r<0 \\
    \scriptstyle
    r \quad 0\le r \le 1 \\ 
    1 \quad r>1
  \end{cases} \quad
  \scriptstyle A_1 =\left[
    \begin{smallmatrix}
      0 & 2 \\[1mm]
      -0.5 & -0.1
    \end{smallmatrix} \right] \quad
  \scriptstyle A_2 =\left[
    \begin{smallmatrix}
      -0.1 & 0.5 \\[1mm]
      -2 & 0
    \end{smallmatrix}\right].
\end{align*}
Note that $g(0,0)=0$ and that $g$ is locally Lipschitz and hence Lipschitz on bounded sets, given the finite dimensionality of the system. This system is a modified version of that in Example~3.5 of \citet{mangar_natma05}, which has the property that solutions exist for all future times when the input is continuous but may have finite escape time for some bounded measurable inputs that are not continuous, as proved in Section~\ref{sec:propNFC}:
\begin{prop}
  \label{prop:NFC}
  System (\ref{eq:planar}) is FCC but not FC.
\end{prop}

As can be easily verified, % the matrices $A_1$ and $A_2$ are both Hurwitz and, moreover, every convex combination 
 $A(\lambda)$ is Hurwitz for every $\lambda \in [0,1]$. Therefore the following lemma holds.
\begin{lem}
  \label{lem:Lyap}
  For each $\lambda\in [0,1]$ there exists a positive definite symmetric matrix $P_{\lambda}\in \R^{2\times 2}$ such that
  \begin{align} \label{eq:Lyap}
    A^T(\lambda)P_{\lambda}+P_{\lambda}A(\lambda)=-I.
  \end{align}
\end{lem}
%% \begin{pf}
%%   The characteristic polynomial of $A(\lambda)$ is $\det (pI - A(\lambda)) = p^2 + 0.1 p + c(\lambda)$, with
%%   $c(\lambda) = -2.26 \lambda^2 + 2.26\lambda + 1$. 
%%   For $\lambda \in [0,1]$, $c(\lambda) \ge 1$, and hence $A(\lambda)$ is Hurwitz. This is equivalent to (\ref{eq:Lyap}) having a unique positive definite solution $P_{\lambda}$ \citep[see, e.g.][Theorem~4.6]{khalil_book02}.\qed
%% \end{pf}
The basis for our main result is the following time-delay system without inputs
\begin{align}
  \label{eq:gasnotugas}
  \dot{z}(t) &= -z(t), &
  \dot{x}(t) &= g(x(t),z(t-\tau)),
  %% \left[\begin{array}{c}
  %%     \dot{z}(t) \\
  %%     \dot{x}(t)
  %%   \end{array} \right] =
  %% \left[\begin{array}{c}
  %%     -z(t) \\
  %%     g(x(t),z(t-\tau))
  %%   \end{array} \right]
\end{align}
where for all $t\ge 0$, $z(t)\in \R$, $x(t)\in \R^2$ and $g$ is as in (\ref{eq:planar}).
\begin{prop}
  \label{prop:main}
  System (\ref{eq:gasnotugas}) is
  \begin{enumerate}[a)]
  \item FC;\label{item:FC0}
  \item ES and UGA; \label{item:LSUAGA0}
%  \item UA,\label{item:UA0}
%  \item GA,\label{item:GA0}
%  \item GAS,\label{item:GAS0}
  \item not RFC.\label{item:NFC0}
%  \item not UGAS.\label{item:NUGAS0}
  \end{enumerate}
\end{prop}
Next, the proof of item~\ref{item:FC0}), which makes use of Proposition~\ref{prop:NFC}, and the rationale for items~\ref{item:LSUAGA0}) and~\ref{item:NFC0}) are provided, for ease of understanding. The precise proof of item~\ref{item:LSUAGA0}) is given in Section~\ref{sec:GAS} and the proof of Proposition~\ref{prop:NFC} in Section~\ref{sec:propNFC}. Thus, the remainder of this section establishes all the positive assertions of Proposition~\ref{prop:main}. The fact that RFC does not hold will be covered in Section~\ref{sec:notRFCnotUGAS}.

\textbf{Proof of \ref{item:FC0}) and rationale of \ref{item:LSUAGA0}) and~\ref{item:NFC0}):}
%The equation for $\dot z$ in~(\ref{eq:gasnotugas}) is decoupled from $x$, and hence the evolution of $z$ for $t\ge 0$ can be readily obtained as the decaying exponential
Given an initial condition $\phi = [z_0,x_0^T]^T \in \C([-\tau,0],\R^3)$, with $z_0\in \C([-\tau,0],\R)$ and $x_0\in  \C([-\tau,0],\R^2)$, from ~(\ref{eq:gasnotugas}) we have that $z(t) = z(0) e^{-t}$ for $t\ge 0$, and that the dynamics of $x$ for $t\in [0,\tau]$ is governed by
%  An initial condition for system~(\ref{eq:gasnotugas}) is of the form $\phi = [z_0,x_0^T]^T \in \C([-\tau,0],\R^3)$, with $z_0\in \C([-\tau,0],\R)$ and $x_0\in  \C([-\tau,0],\R^2)$. Given an initial condition, the dynamics of $x$ for $t\in [0,\tau]$ is governed by
\begin{align}
  \label{eq:NFCxini}
  \dot x(t) = g\big(x(t),\underbrace{z_0(t-\tau)}_{v(t)}\big),
  \quad t\in [0, \tau],
\end{align}
where $z_0(t-\tau) = z(t-\tau)$ can be interpreted as a (continuous) input $v(t)$. %The system evolution for $t\in [0,\tau]$ is thus governed by the equations of a nondelayed second-order, and decoupled system with an external continuous input.
 For $t\ge \tau$, the dynamics of $x$ becomes 
\begin{align*}
  \dot x(t) &= g(x(t), \underbrace{z_0(0) e^{-(t-\tau)}}_{w(t)}),\quad t\ge\tau,
\end{align*}
which can also be regarded as a nondelayed system with the external continuous input $w$. Since $v(\tau) = w(\tau)$, then the evolution of $x$ for all $t\ge 0$ can be seen to be governed by the system~(\ref{eq:planar}) with the continuous input $u$ being the concatenation of $v$ and $w$ at instant $\tau$. By Proposition~\ref{prop:NFC} solutions of the time-delay system~(\ref{eq:gasnotugas}) must exist for all future times and thus~(\ref{eq:gasnotugas}) is FC. 

The rationale for the remaining items is as follows. 

Item~\ref{item:LSUAGA0}). When the initial condition has small norm, the system~(\ref{eq:planar}) becomes close to its linearization at the origin corresponding to zero input and $x$ behaves almost as a stable linear system. This is the rationale for ES.
The fact that~(\ref{eq:gasnotugas}) is UGA can be conceptually explained as follows. The factor $(1+|x(t)|^2)$ in~(\ref{eq:planar}) causes the state evolution of~(\ref{eq:gasnotugas}) to both grow and decrease very rapidly over compact time intervals, but in a way such that the reachability times from bounded balls to bounded balls are uniformly bounded. Therefore, a bounded reachability time is maintained.

Item~\ref{item:NFC0}). Example 3.5 in \citet{mangar_natma05} also shows that system~(\ref{eq:planar}) is not FC and therefore if the input $u$ is not continuous, solutions may fail to exist for all future times. By virtue of the BIC property, a solution can only cease to exist if it has a finite escape time. The rationale of the proof that RFC does not hold is then to take an equibounded sequence of continuous functions, each of which serving as an initial condition $z_0$, so that this sequence approaches a bounded measurable discontinuous function for which system~(\ref{eq:planar}) has a finite escape time. The (bounded) state evolutions corresponding to each of these initial conditions will have bounds that grow larger as the limiting discontinuous function is approached, and over a compact time interval. Since no uniform bound exists for these solutions over a compact time interval, then RFC does not hold. Note that even if the trajectories corresponding to the aforementioned sequence of initial conditions have larger and larger intermediate peaks, uniform reachability times are maintained for UGA to be possible. This seems to be counterintuitive. 
\qed

Note that~(\ref{eq:gasnotugas}), being ES (and therefore LS) and UGA according to Proposition~\ref{prop:main}, is hence GAS. If (\ref{eq:gasnotugas}) is not RFC, and since UGAS is equivalent to the combination of GAS and RFC \citep[Theorem~1]{karpep_jco22}, it follows that (\ref{eq:gasnotugas}) cannot be UGAS, either. These comments are the proof of the following.
\begin{cor}
  \label{cor:uga+gas-ugas}
  System~(\ref{eq:gasnotugas}) is UGA+GAS but not UGAS.
\end{cor}
%The remainder of this section contains the proof of Proposition~\ref{prop:main}\ref{item:LSUAGA0}) in Section~\ref{sec:GAS} and the proof of Proposition~\ref{prop:NFC} in Section~\ref{sec:propNFC}. This section thus proves all the positive assertions of Proposition~\ref{prop:main}. The fact that RFC does not hold will be covered in Section~\ref{sec:notRFCnotUGAS}.

\subsection{Lyapunov Stability and Attractivity}
\label{sec:GAS}
  
Next, ES and UGA of~(\ref{eq:gasnotugas}) are established.
%{\bf Proof of Proposition~\ref{prop:main}\ref{item:GAS0})}. Consider the non-delayed system with inputs 
%\begin{align}\label{eq:syst1}
%\left [\begin{array}{c} \dot{z}(t) \\
%\dot{x}(t) \end{array} \right ] = \left [\begin{array}{c} -z(t) \\
%g(x(t),u(t))\end{array} \right ]
%\end{align}
%where for $t\ge 0$, $u(t)\in \R$.
By Lemma~\ref{lem:Lyap}, there exists a symmetric positive definite matrix $P_0\in \R^{2\times 2}$ such that $A(0)^TP_0+P_0A(0)=-I$. By continuity, there exists $\Lambda\in (0,1)$ such that $A(\lambda)^TP_0+P_0A(\lambda)\le -\frac{1}{2}I$ for all $0\le \lambda \le \Lambda$. Define $W:\R^2\to \R_{\ge 0}$, $W(x)=x^TP_0x$ for all $x \in \R^2$. There exist positive $c_1,c_2$ so that
\begin{align} \label{eq:cis}
c_1|x|_2^2 \le W(x) \le c_2 |x|_{2}^2\quad \forall x\in \R^2.
\end{align}
%
%Pick any $0<r\le\Lambda$.
Let $[z(t)\;x(t)^T]^T$ be the solution of (\ref{eq:gasnotugas}) corresponding to an initial condition $[z_0,x_0^T]^T \in \C([-\tau,0],\R^3)$ such that $\|z_0\|\le \Lambda$ and $\|x_0\|\le \Lambda$.
Then
\begin{align}
  \label{eq:ztexp}
  |z(t)| &=|z(0)|e^{-t}\le \|z_0\|e^{-t} \text{ for all }t\ge 0.
\end{align}
Also, $0\le \varphi(z(t-\tau))\le |z(t-\tau)|\le \|z_0\|\le \Lambda$ for all $t\ge 0$. For $\omega(t):=W(x(t))$ and $\zeta(t) := \varphi(z(t-\tau))$, it follows that
\begin{align} 
  &\dot{\omega}(t) = (1+|x(t)|_{2}^2) x^T(t) \left[A^T(\zeta(t)) P_0+P_0A(\zeta(t))\right] x(t)\notag\\
  \label{eq:dec-gen}
  &\le -(1+|x(t)|_{2}^2)\frac{|x(t)|_2^2}{2} \le -\frac{\omega(t)}{2c_2} -\frac{\omega^2(t)}{2c_2^2} \le 0.
  %% \\
  %% \label{eq:dec}
  %% &\le -\frac{|x(t)|_2^2}{2} \le -\frac{w(t)}{2c_2}.
\end{align}
Then, $\dot\omega(t)\le -\frac{\omega(t)}{2c_2}$ for all $t\ge 0$ and hence, by Gronwall's lemma, $\omega(t)\le \omega(0)e^{-\frac{t}{2c_2}}$ for all $t\ge 0$. Hence $|x(t)|_2\le \sqrt{c_2/c_1}|x(0)|_2 e^{-\frac{t}{4c_2}}$ and thus $|x(t)|\le \sqrt{2c_2/c_1}|x(0)| e^{-\frac{t}{4c_2}}$. Therefore $| [z(t),\;x(t)^T]^T|_{\infty} \le k \|\phi\|e^{-p t}$ for all $t\ge 0$, with $k=\sqrt{2c_2/c_1}$ and $p=\min\{1,\frac{1}{4c_2}\}$, happens if $\|[z_0,x_0^T]^T\|\le\Lambda$, showing that~(\ref{eq:gasnotugas}) is ES.

To establish UGA, let $r$ and $\varepsilon$ be given, consider an initial condition $\phi_0 = [z_0,x_0^T]^T \in B^{\mathcal{C}}_{r}$ and let $[z(t),x(t)^T]^T$ be the corresponding solution of~(\ref{eq:gasnotugas}). From~(\ref{eq:ztexp}), there exists $t_1 = t_1(r,\varepsilon)$ such that $|z(t)| \le \min\{\Lambda,\varepsilon\}$ for all $t\ge t_1$, with $\Lambda$ as above. Defining $\omega(t) = W(x(t))$ as above, then~(\ref{eq:dec-gen}) holds for all $t\ge t_1 +\tau$, which implies that
%% From~(\ref{eq:dec-gen}) and~(\ref{eq:cis}),
%% \begin{align*}
%%   \dot w(t) &\le -\frac{w^2(t)}{2c_2^2},\quad t\ge t_1 + \tau,
%% \end{align*}
$\dot{\omega}(t)\le -\frac{\omega^2(t)}{2c_2^2}$ for all $t\ge t_1+\tau$. By solving $\dot{y}(t)=-\frac{y^2(t)}{2c_2^2}$, $y(t_1+\tau)=\omega(t_1+\tau)$ and using a comparison lemma, then $\omega(t)\le y(t)$ for all $t\ge t_1+\tau$, \emph{i.e.}
\begin{align*}
  \omega(t) \le \frac{2c_2^2 \omega(t_1+\tau)}{2c_2^2+\omega(t_1+\tau)[t-(t_1+\tau)]} \le \frac{2c_2^2}{t-(t_1+\tau)}
\end{align*}
for $t> t_1 + \tau$. Then, there exists $T = T(r,\varepsilon) \ge t_1 + \tau$, so that $\omega(t) \le c_1 \varepsilon^2$ for all $t\ge T$, and hence $|x(t)| \le |x(t)|_2 \le \varepsilon$ for all $t\ge T$. This establishes UGA because $\|[z(t),x(t)^T]\|\le \varepsilon$ for all $t\ge T$.
\qed
%% Regarding global attractivity, let $[z(t)\;x(t)^T]^T$ be the solution of (\ref{eq:gasnotugas}) corresponding to an initial condition $[z_0,x_0^T]^T$. Since $|z(t)|\le |z(0)|e^{-t}$ for all $t\ge 0$, there exists $T>0$ such that $|z(t)|<\Lambda$ for all $t\ge T$, with $\Lambda$ as above. Then $w(t)=W(x(t))$ satisfies (\ref{eq:dec}) for all $t\ge T+\tau$ and therefore, $w(t)\to 0$ as $t\to \infty$, which in turn implies that $x(t)\to 0$ as $t\to \infty$. In consequence $[z(t)\;x(t)^T]^T \to 0$ as $t\to \infty$ and GA is established. \qed

\subsection{Proof of Proposition~\ref{prop:NFC}}
\label{sec:propNFC}

%Proposition~\ref{prop:NFC} shows that there is an important gap between the behavior of solutions corresponding to Lebesgue measurable locally essentially bounded inputs on the one hand and continuous inputs on the other. This proposition is needed to establish item~\ref{item:FC0}) of Proposition~\ref{prop:main}, namely that system~(\ref{eq:gasnotugas}) is FC.

%{\bf Proof of Proposition \ref{prop:NFC}}.
Example 3.5 in \citet{mangar_natma05} proves that system (\ref{eq:planar}) is not FC.

Next, it will be shown that (\ref{eq:planar}) is FCC.
Let $\xi_0\in \R^2$ and let $u:\R_{\ge 0}\to \R$ be continuous. Let $x(\cdot)=x(\cdot,\xi_0,u)$ be the maximally defined solution of (\ref{eq:planar}). Let $[0,T)$ with $T>0$ be the interval of definition of $x(\cdot)$ and, for a contradiction, suppose that $T<\infty$. Then, due to the BIC property, $x(\cdot)$ is unbounded on $[0,T)$. 
%
%Let $\lambda^*=\varphi(u(T))$ and 
Let $P_{\lambda^*}$ be the positive definite symmetric matrix from Lemma~\ref{lem:Lyap} corresponding to $\lambda^*=\varphi(u(T))$.
%that satisfies~(\ref{eq:Lyap}) with $\lambda^*$ instead of $\lambda$. 
By continuity, there exists $\delta^*>0$ such that $A^T(\lambda)P_{\lambda^*}+P_{\lambda^*}A(\lambda)\le -\frac{1}{2}I$ for all $\lambda \in (\lambda^*-\delta^*, \lambda^*+\delta^*)$. Since $\varphi(u(\cdot))$ is continuous, there exists $\delta>0$ such that $|\varphi(u(t))-\varphi(u(T))|=|\varphi(u(t))-\lambda^*|<\delta^*$ for all $t \in (T-\delta,T+\delta)$. 
Define $W(x)=x^TP_{\lambda^*}x$ for all $x\in \R^2$. Then, there exist positive $c_1,c_2$ such that (\ref{eq:cis}) holds.
For $t\in (T-\delta,T)$, define $\omega(t)=W(x(t))$. It follows that $\omega(\cdot)$ is unbounded since $x(\cdot)$ is unbounded on $(T-\delta,T)$.
% and $\omega(t)\ge c_1|x(t)|_{2}^2$ for all $t\in (T-\delta,T)$. 
However, similarly to~(\ref{eq:dec-gen}), one finds that $\dot{\omega}(t)\le 0$ for $t\in (T-\delta,T)$.
%\begin{align*}
%  \dot{\omega}(t) %&=\left (1+|x(t)|_2^2 \right )\cdot\\
  %&\phantom{=(}\left[x(t)^T\Big(A^T(\varphi(u(t)))P_{\lambda^*}+P_{\lambda^*}A(\varphi(u(t))\Big) x(t)\right]\\
%  &\le -\left (1+|x(t)|_2^2 \right ) \frac{|x(t)|_{2}^2}{2}\le 0%\end{align*}
Therefore $\omega(\cdot)$ is nonincreasing on $(T-\delta,T)$ and hence, since it is also nonnegative, must be bounded, reaching a contradiction. So, the interval of definition of $x(\cdot)$ is $[0,\infty)$ and (\ref{eq:planar}) is FCC.

%{\bf Proof of Proposition~\ref{prop:main}\ref{item:FC0}):} As explained in Section~\ref{sec:example-rationale}, for every initial condition $\phi = [z_0,x_0^T]^T$, the resulting system evolution for $t\ge 0$ is given by $[z(t), x(t)^T]^T$ with $z(t) = z_0(0) e^{-t}$ and $x(t)$ the solution to system~(\ref{eq:planar}) under a continuous input. By Proposition~\ref{prop:NFC}, system~(\ref{eq:planar}) is FC for continuous inputs, and hence the solution of~(\ref{eq:gasnotugas}) exists for all $t\ge 0$. \qed

\section{Characterizations of FC and RFC}
\label{sec:notRFCnotUGAS}

The proof that system~(\ref{eq:gasnotugas}) is not RFC is based on a characterization of RFC for time-delay systems whose evolution is governed equivalently by nondelayed systems with continuous inputs. This characterization is obtained next.

Consider a time-delay system of the form
\begin{align}\label{eq:sistrp}
\dot{x}(t)=\mathsf{f}\Big(x(t),\big(x(t-\tau_1),\ldots,x(t-\tau_{\ell})\big),u(t)\Big),
\end{align}
where $\mathsf{f}:\R^n \times \R^{\ell n} \times \R^m\to \R^n$ is locally Lipschitz, and 
\begin{subequations}
  \label{eq:taus}
  \begin{align}
    \tau_0 &:= 0 < \tau_1 <\cdots <\tau_{\ell} =: \tau\\
    \tau^* &=\min\{\tau_{i}-\tau_{i-1}:i=1,\ldots,\ell\}.
  \end{align}
\end{subequations}
System (\ref{eq:sistrp}) can be written as in (\ref{eq:sistr}) with 
$$f(\phi,\mu)=\mathsf{f}(\phi(0),(\phi(-\tau_1),\ldots,\phi(-\tau_{\ell})),\mu)$$ for all $\phi \in \C$ and $\mu \in \R^m$, with $f$ satisfying Assumption~\ref{as:Lipf} due to the assumption on $\mathsf{f}$. Associated with~(\ref{eq:sistrp}), consider the nondelayed system with $\ell n+m$ inputs
\begin{align}
  \label{eq:sistnd}
  \dot{z}(t) &=\mathsf{f}(z(t),\underbrace{(v_1(t),\ldots,v_{\ell}(t))}_{\mathsf{v}(t)},u(t))
  %% \notag\\
  %% &=\mathsf{f}(z(t),\mathsf{v}(t),u(t))
\end{align}
 where $v_i\in \U^n$ for $i=1,\ldots,\ell$, $\mathsf{v}=(v_1,\ldots,v_{\ell})\in \U^{\ell n}$ and $u\in \U^m$. We will denote by $z(\cdot,\xi_0,\mathsf{v},u)$ the maximal solution of (\ref{eq:sistnd}) corresponding to the inputs $\mathsf{v}\in \U^{\ell n}$ and $u\in \U^m$ such that $z(0)=\xi_0\in \R^n$.

The following result states that RFC of the time-delay system~(\ref{eq:sistrp}) is equivalent to FC of the nondelayed system~(\ref{eq:sistnd}). The key point here is that the latter being FC means that solutions of~(\ref{eq:sistnd}) should exist for all future times \emph{for all Lebesgue measurable and locally essentially bounded inputs}, and not just for continuous inputs.
\begin{thm}
  \label{thm:RFCchar}
  The following statements are equivalent.
  \begin{enumerate}[a)]
  \item \label{item:RFC} System (\ref{eq:sistrp}) is RFC.
  \item \label{item:FC} System (\ref{eq:sistnd}) is FC.
  %% \item \label{item:LF} There exists a proper and smooth function $V:\R^n\to \R_{\ge 0}$ such that
  %%   \begin{align}
  %%     \nabla V(x)\mathsf{f}(\xi,\nu,\mu)\le V(\xi)+\sigma_1(|\nu|)+\sigma_2(|\mu|)\quad \forall \xi\in \R^n,\forall\nu \in \in \R^{\ell n},\forall \mu \in \R^m, 
  %%   \end{align}
  %%   holds for some $\sigma_1,\sigma_{2}\in \Ki$.
  \end{enumerate}
\end{thm}
{\bf Proof.} Given $T\ge 0$ and $r\ge 0$, define
\begin{align}
  R^*(r,&T) = \sup \{|z(t,\xi_0,\mathsf{v},u)|:\notag\\
  \label{eq:reachable}
  &|\xi_0|\le r, \|\mathsf{v}\|_{\infty}\le r, \|u\|_{\infty}\le r, 0\le t\le T\},\\
  R(r,&T) =\sup \{|x(t,\phi,u)|:\notag\\
  \label{eq:reachabletd}
  &\hspace{10mm}\|\phi\|\le r, \|u\|_{\infty}\le r, -\tau\le t\le T\},
\end{align}
where $x(\cdot,\phi,u)$ stands for the maximal solution of (\ref{eq:sistrp}) corresponding to the initial condition $\phi \in \mathcal{C}$ and the input $u\in \U^m$. Both $R$ and $R^*$ are nondecreasing in each of its variables, and finite for every $(r,T)\in \R^2_{\ge 0}$ if and only if the corresponding system is $RFC$. Due to time-invariance and the semigroup property of the solutions, it is also true that $R(r,t_1+t_2)\le R(R(r,t_1),t_2)$ and $R^*(r,t_1+t_2)\le R^*(R^*(r,t_1),t_2)$ for every $r\ge 0$ and $t_1,t_2\ge 0$.
Therefore, if for some $\bar \tau>0$ it happens that $R(r,\bar \tau)$ is finite for all $r\ge 0$ then $R(r,T)$ is finite for all $(r,T)\in \R^2_{\ge 0}$ and hence (\ref{eq:sistrp}) is RFC. The same holds for (\ref{eq:sistnd}) if $R^*(r,\bar \tau)$ is finite for all $r\ge 0$.
Recall~(\ref{eq:taus}). % and let $\tau^*=\min\{\tau_{i}-\tau_{i-1}:i=1,\ldots,\ell\}$.

\ref{item:FC}) $\Rightarrow$ \ref{item:RFC}). Suppose that the nondelayed system~(\ref{eq:sistnd}) is FC. By Proposition~5.1 of \citet{linson_jco96} it follows that (\ref{eq:sistnd}) is RFC.
Let $r\ge 0$, $\phi\in \mathcal{C}$ and $u\in \U^m$ be such that $\|\phi\|\le r$ and $\|u\|_{\infty}\le r$. Define for $i=1,\ldots,\ell$, $v_i(t)=\phi(t-\tau_i)$ if $t\in [0,\tau_1)$ and $v_i(t)=0$ elsewhere. Set $\xi_0=\phi(0)$. Then, if $x(t)=x(t,\phi,u)$ and $z(t)=z(t,\xi_0,\mathsf{v},u)$, we have that for all $0\le t <\min\{\tau_1,t_{(\phi,u)}\}$
\begin{align*}
x(t)&=\phi(0)+\int_{0}^t\mathsf{f}\big(x(s),{\scriptstyle (x(s-\tau_1),\ldots,x(s-\tau_{\ell}))},u(s)\big)\:ds \\
%&=\xi_0+\int_{0}^t\mathsf{f}\big(x(s),{\scriptstyle (\phi(s-\tau_1),\ldots,\phi(s-\tau_{\ell}))},u(s)\big)\:ds \\
&=\xi_0+\int_{0}^t\mathsf{f}\big(x(s),(v_1(s),\ldots,v_{\ell}(s)),u(s)\big)\:ds 
\end{align*}
Thus $x(\cdot)$ restricted to $[0,\min\{\tau_1,t_{(\phi,u)}\})$  is a solution  of (\ref{eq:sistnd}) such that $x(0)=\xi_0$, and therefore $z(t)=x(t)$ for all $t\in [0,\min\{\tau_1,t_{(\phi,u)}\})$. Since $|\xi_0|\le r$, $\|\mathsf{v}\|_{\infty}\le r$ and $\|u\|_{\infty}\le r$, we have that $|z(t)|\le R^*(r,\tau_1)$ and, in consequence, $t_{(\phi,u)}> \tau_1$ due to the BIC property of (\ref{eq:sistrp}). By continuity, $z(t)=x(t)$ for all $t\in [0,\tau_1]$ and therefore $|x(t)|\le R^*(r,\tau_1)$ for all $t\in [0,\tau_1]$. Taking into account that $R^*(r,\tau_1)\ge R^*(r,0)=r$ and that $|x(s)|=|\phi(s)|\le r$ for all $s\in [-\tau,0]$, it then follows that $|x(t)|\le R^*(r,\tau_1)$ for all $t\in [-\tau,\tau_1]$. We have then proved that $R(r,\tau_1)\le R^*(r,\tau_1)$ for all $r\ge 0$ and then that (\ref{eq:sistrp}) is RFC.

\ref{item:RFC}) $\Rightarrow$ \ref{item:FC}). Suppose that the time-delay system~(\ref{eq:sistrp}) is RFC. Let $\tau^*$ be as in~(\ref{eq:taus}). Let $r\ge 0$ and $z(t)=z(t,\xi_0,\mathsf{v},u)$ with $|\xi_0|\le r$, $\|\mathsf{v}\|_{\infty}\le r$ and $\|u\|_{\infty}\le r$. Denote the maximum time of existence of $z$ by $t_{z}$.
For each component $v_i$ of $\mathsf{v}$ there exists a sequence of continuous functions $\{w_{i,n}\}_{n\in\N}$, $w_{i,n} \in \C([0,\tau_i-\tau_{i-1}], \R^n)$ such that $\|w_{i,n}\|_{\infty}\le r$ and for almost all $t\in [0,\tau_i-\tau_{i-1}]$, $\lim_{n\to\infty} w_{i,n}(t) = v_i(t)$ (pointwise almost everywhere convergence). We can assume without loss of generality that $w_{1,n}(\tau_1)=\xi_0$ and that $w_{i,n}(\tau_i-\tau_{i-1})=w_{i-1,n}(0)$ for all $2\le i \le \ell$ and for all $n\in\N$. Define $\phi_n:[-\tau,0]\to \R^n$ via
$\phi_n(s)=w_{i,n}(s+\tau_i)$ for all $s\in [-\tau_i,-\tau_{i-1})$, for all $i=1,\ldots,\ell$ and $\phi_n(0)=\xi_0$. We have that $\phi_n\in \C$ and $\|\phi\|\le r$.

Let $x_n(t)=x(t,\phi_n,u)$. Since~(\ref{eq:sistrp}) is RFC, we have that $|x_n(t)|\le R(r,\tau^*)<\infty$ for all $0\le t\le \tau^*$ and all $n\in\N$. We also have that for all $t\in [0,\tau^*]$
\begin{align*}
x_n(t)&=\xi_0+\int_{0}^t \mathsf{f}\big(x_n(s),{\scriptstyle (x_n(s-\tau_1), \ldots, x_n(s-\tau_\ell))}, u(s) \big) \: ds\displaybreak[0]\\
%&=\xi_0+\int_{0}^t \mathsf{f}\big(x_n(s),{\scriptstyle (\phi_n(s-\tau_1),\ldots, \phi_n(s-\tau_\ell))},u(s)\big) \:ds\\
&=\xi_0+\int_{0}^t \mathsf{f}\big(x_n(s),{\scriptstyle (w_{1,n}(s),\ldots, w_{\ell,n}(s))},u(s)\big)\:ds.
\end{align*}
So, $x_n(t)=z_n(t)=z(t,\xi_0,\mathsf{w}_n,u)$ for all $t\in [0,\tau^*]$, where $\mathsf{w_n}$ is the input whose $i$-th block of $n$ components, restricted to $[0,\tau^*]$, is the function $w_{i,n}$ previously defined and that is zero elsewhere.
Since the sequence $\{x_n\}$ is uniformly bounded on $[0,\tau^*]$ and $\mathsf{f}$ is bounded on bounded sets, it follows that there exists a constant $M$ such that $$|\dot{x}_n(t)|=|\mathsf{f}(x_n(t),(w_{1,n}(t),\ldots, w_{\ell,n}(t)),u(t))|\le M$$ for almost all $t\in [0,\tau^*]$ and all $n\in\N$. Hence, by the Arzel\`a-Ascoli Theorem, there exists a subsequence $\{x_{n_k}\}$ which converges uniformly to a continuous function $x:[0,\tau^*]\to \R^n$ on $[0,\tau^*]$. Since $\mathsf{f}$ is continuous, $\mathsf{w_n}\to \mathsf{v}$ a.e. on $[0,\tau^*]$ and $|\mathsf{f}(x_n(s),(w_{1,n}(s),\ldots, w_{\ell,n}(s)),u(s))|\le M$ for almost all $s\in [0,\tau^*]$, by the Lebesgue Dominated Convergence Theorem it follows that
\begin{align*}
&x(t)=\lim_{k\to \infty}x_{n_k}(t)\\\displaybreak[0]
&= \xi_0+\lim_{k\to \infty}\int_{0}^t \mathsf{f}(x_{n_k}(s), {\scriptstyle (w_{1,n_k}(s),\ldots, w_{\ell,n_k}(s))},u(s))\:ds\\
&=\xi_0+\int_0^t \mathsf{f}(x(s),\mathsf{v}(s),u(s))\:ds\quad \forall t\in [0,\tau^*].
\end{align*}
In consequence, $z(t)=x(t)$ and $|z(t)|\le R(r,\tau^*)$ for all $t\in [0,\tau^*]$. This establishes that $R^*(r,\tau^*)\le R(r,\tau^*)<\infty$ and {\em a fortiori} that (\ref{eq:sistnd}) is FC.
%
%% The equivalence between \ref{item:FC}) and \ref{item:LF}) is a straighforward consequence of Corollary 2.11 in \cite{angson_scl99}.
\qed

{\bf Proof of Proposition~\ref{prop:main}\ref{item:NFC0}):}
From Proposition \ref{prop:NFC} we have that (\ref{eq:planar}) is not FC, and hence the system
\begin{align}
  \label{eq:syst1}
  \dot{z}(t) &= -z(t), &\dot{x}(t) &= g(x(t),u(t)),
%% \left [\begin{array}{c} \dot{z}(t) \\
%% \dot{x}(t) \end{array} \right ] = \left [\begin{array}{c} -z(t) \\
%% g(x(t),u(t))\end{array} \right ]
\end{align}
where $u\in \U^1$, is not FC. By Theorem~\ref{thm:RFCchar}, it follows that (\ref{eq:gasnotugas}) cannot be RFC.
%Since system~(\ref{eq:gasnotugas}) is not RFC, it cannot be UGAS by Theorem 1 of \citet{karpep_jco22}. 
\qed

The following characterization
%of the FC of (\ref{eq:sistrp}) in terms of the FC of system (\ref{eq:sistnd}) when the inputs $\mathsf{v}$ are continuous
is a %straightforward 
consequence of the relationship between initial conditions of (\ref{eq:sistrp}) and continuous inputs $\mathsf{v}$ of (\ref{eq:sistnd}). We say that (\ref{eq:sistnd}) is FC for inputs $(\mathsf{v},u)$ in $\U_c^{\ell n} \times \U^m$ if for all $\xi_0\in \R^n$ and  $(\mathsf{v},u)\in \U_c^{\ell n} \times \U^m$, $z(t,\xi_0,\mathsf{v},u)$ is defined for all $t\ge 0$.
\begin{thm} \label{thm:FCchar} The following statements are equivalent.
\begin{enumerate}[a)]
\item \label{item:FCtd} System (\ref{eq:sistrp}) is FC.
\item \label{item:FCc} System (\ref{eq:sistnd}) is FC for inputs $(\mathsf{v},u)$ in $\U_c^{\ell n} \times \U^m$.
\end{enumerate}
\end{thm}
\begin{pf}
  \ref{item:FCc}) $\Rightarrow$ \ref{item:FCtd}). Let $\phi \in \C$, $u_0\in\U^m$. Induction will be used to establish that $x(t,\phi,u_0)$ is defined for all $t\ge 0$. Define $v_i(t) = \phi(t-\tau_i)$ for $t\in [0,\tau_1]$, $v_i(t) = \phi(\tau_1-\tau_i)$ for $t>\tau_1$, $i=1,\ldots\ell$, and $\xi_0=\phi(0)$, so that $v_i \in \U^n_c$ and $z(t,\xi_0,\mathsf{v},u_0) = x(t,\phi,u_0)$ for $t\in [0,\tau_1] \cap [0,t_{(\phi,u_0)})$. By~\ref{item:FCc}) and the BIC property, then $t_{(\phi,u_0)} > \tau_1$ and hence $x(t,\phi,u_0)$ exists for $t\in [0,\tau_1]$. Next, suppose that $t_{(\phi,u_0)}>k\tau_1$ for some $k\in\N$. Let $\phi_k=x_{k\tau_1}$ and $u_k=u_0(\cdot-k\tau_1)$. By using the same arguments as in the first part, we have that $x(t,\phi_k,u_k)$ exists for all $t\in[0,\tau_1]$. Since $x(t,\phi,u_0)=x(t-k\tau_1,\phi_k,u_k)$ for all $t\in [k\tau_1,t_{(\phi,u_0)})$ it follows that $t_{(\phi,u_0)}>(k+1)\tau_1$ and the proof is complete.
% Define $v_i(t) = x(t+k\tau_1-\tau_i,\phi,u_0)$ for $t\in [0,\tau_1]$, $v_i(t) = x((k+1)\tau_1-\tau_i,\phi,u_0)$ for $t>\tau_1$, $i=1,\ldots,\ell$, $\xi_k=x(k\tau_1,\phi,u_0)$, $u_k(\cdot)=u_0(\cdot + k\tau_1)$, so that $v_i\in\U^n_c$ and $z(t,\xi_k,\mathsf{v},u_k) = x(t+k\tau_1,\phi,u_0)$ for $t\in [0,\tau_1] \cap [0,t_{(\phi,u_0)} - k\tau_1)$. By~\ref{item:FCc}), the BIC property and the induction assumption, then $x(t,\phi,u_0)$ exists for $t\in [0,(k+1)\tau_1]$ and the proof by induction is complete.

  \ref{item:FCtd}) $\Rightarrow$ \ref{item:FCc}). Let $\xi_0 \in \R^n$, $\mathsf{v} \in \U_c^{\ell n}$, $u\in\U^m$ and recall~(\ref{eq:taus}). Define $\phi(t) = v_i(t+\tau_i)$ for $t\in [-\tau_i,-\tau_i+\frac{\tau^*}{2}]$, $i=1,\ldots,\ell$, and $\phi(0) = \xi_0$. Complete $\phi$ into a continuous function defined on $[-\tau,0]$ (\emph{e.g.} via linear interpolation). %i.e. $\phi(t) = \sigma(t) v_i(\tau^*/2) + [1-\sigma(t)] v_{i-1}(\tau_{i-1})$ for $t\in (-\tau_i+\frac{\tau^*}{2},-\tau_{i-1})$, with $\sigma(t) = \dfrac{-\tau_{i-1} - t}{\tau_i-\tau_{i-1} - \tau^*/2}$, for $i=1,\ldots,\ell$ and $v_0(\tau_0) := \phi(0)$. 
  Then, $\phi\in\C$ and $z(t,\xi_0,\mathsf{v},u) = x(t,\phi,u)$ for $t\in[0,\tau^*/2] \cap [0,t_{(\xi_0,\mathsf{v},u)})$ with $t_{(\xi_0,\mathsf{v},u)}$ the maximal time of existence of $z$. By \ref{item:FCtd}) and BIC, then $t_{(\xi_0,\mathsf{v},u)} > \tau^*/2$. Next, suppose that $t>k\tau^*/2$ for some $k\in\N$.
Define $\xi_k=z(k\tau^*/2,\xi_0,\mathsf{v},u)$, $\mathsf{v}_k=\mathsf{v}(\cdot-k\tau^*/2)$ and $u_k=u(\cdot-k\tau^*/2)$. With the same arguments as above, $z(t,\xi_k,\mathsf{v}_k,u_k)$ exists for all $t\in[0,\tau^*/2]$. Since $z(t,\xi_0,\mathsf{v},u)=z(t-k\tau^*/2,\xi_k,\mathsf{v}_k,u_k)$ for all $t\in [k\tau^*/2, t_{(\xi_0,\mathsf{v},u)})$, it follows that $t_{(\xi_0,\mathsf{v},u)}>(k+1)\tau^*/2$. By induction, then $t_{(\xi_0,\mathsf{v},u)} = \infty$ and \ref{item:FCc}) is established.
  %$z$ is defined for $t\in [0,k\tau^*/2]$ for some $k\in\N$. Define $\phi(t) = v_i(t+\tau_i+k\tau^*/2)$ for $t\in [-\tau_i,-\tau_i+\frac{\tau^*}{2}]$, $i=1,\ldots,\ell$, and $\phi(0) = z(k\tau^*/2,\xi_0,\mathsf{v},u)$. Again, complete $\phi$ into a continuous function via linear interpolation, so that $\phi\in\C$. By \ref{item:FCtd}), BIC, and the induction assumption, then $t_{(\xi_0,\mathsf{v},u)} > (k+1)\tau^*/2$. By induction, then $t_{(\xi_0,\mathsf{v},u)} = \infty$ and \ref{item:FCc}) is established.
\end{pf}

\section{Conclusions}

It was shown that forward completeness is not equivalent to robust forward completeness for time-delay systems, by means of a counterexample. This counterexample, in addition to being forward complete, is exponentially stable and uniformly globally attractive. This establishes the perhaps highly counterintuitive fact that not even the combination of the latter three properties suffices to imply robust forward completeness. It was also shown that the robust forward completeness of %some usually encountered classes of time-delay systems
systems having a finite number of discrete delays
is equivalent to the forward completeness of an associated finite-dimensional nondelayed system under Lebesgue measurable and locally essentially bounded inputs. The existence of nondelayed systems that are forward complete for continuous inputs but not in general for measurable inputs then prevents forward completeness to become equivalent to robust forward completeness. A corollary of the given results is the fact that global asymptotic stability is not equivalent to uniform global asymptotic stability for time-delay systems.
%, giving a negative answer to another current conjecture.

\bibliographystyle{plainnat}
%% \bibliographystyle{plain}
%\bibliography{strings.bib,complete_v2.bib,hernan_v2.bib}
\bibliography{/home/hhaimo/latex/strings.bib,/home/hhaimo/latex/complete_v2.bib,/home/hhaimo/latex/Publications/hernan_v2.bib}

\begin{thebibliography}{9}
\providecommand{\natexlab}[1]{#1}
\providecommand{\url}[1]{\texttt{#1}}
\expandafter\ifx\csname urlstyle\endcsname\relax
  \providecommand{\doi}[1]{doi: #1}\else
  \providecommand{\doi}{doi: \begingroup \urlstyle{rm}\Url}\fi

\bibitem[Chaillet et~al.(2023)Chaillet, Karafyllis, Pepe, and
  Wang]{chakar_mcss23}
A.~Chaillet, I.~Karafyllis, P.~Pepe, and Y.~Wang.
\newblock The {ISS} framework for time-delay systems: a survey.
\newblock \emph{Mathematics of Control, Signals and Systems}, 35:\penalty0
  237--306, 2023.

\bibitem[Hahn(1967)]{hahn_book67}
W.~Hahn.
\newblock \emph{Stability of motion}.
\newblock Springer-Verlag, 1967.

\bibitem[Hale and Lunel(1993)]{halver_book93}
J.~K. Hale and S.~M.~Verduyn Lunel.
\newblock \emph{Introduction to functional differential equations}.
\newblock Springer-Verlag New York, 1993.

\bibitem[Karafyllis and Jiang(2011)]{karjia_book11}
I.~Karafyllis and Z.-P. Jiang.
\newblock \emph{Stability and stabilization of nonlinear systems}.
\newblock Springer London, 2011.

\bibitem[Karafyllis et~al.(2022)Karafyllis, Pepe, Chaillet, and
  Wang]{karpep_jco22}
I.~Karafyllis, P.~Pepe, A.~Chaillet, and Y.~Wang.
\newblock Is global asymptotic stability necessarily uniform for time-invariant
  time-delay systems?
\newblock \emph{SIAM J. Control and Optimization}, 60\penalty0 (6):\penalty0
  3237--3261, 2022.

\bibitem[Lin et~al.(1996)Lin, Sontag, and Wang]{linson_jco96}
Y.~Lin, E.~D. Sontag, and Y.~Wang.
\newblock A smooth converse {L}yapunov theorem for robust stability.
\newblock \emph{SIAM J. Control and Optimization}, 34\penalty0 (1):\penalty0
  124--160, 1996.

\bibitem[Mancilla-Aguilar et~al.(2005)Mancilla-Aguilar, Garc\'{\i}a, Sontag,
  and Wang]{mangar_natma05}
J.~L. Mancilla-Aguilar, R.~A. Garc\'{\i}a, E.~D. Sontag, and Y.~Wang.
\newblock On the representation of switched systems with inputs by perturbed
  control systems.
\newblock \emph{Nonlinear Analysis: Theory, Methods \& Applications},
  60\penalty0 (6):\penalty0 1111--1150, 2005.

\bibitem[Massera(1956)]{masser_am56}
J.~L. Massera.
\newblock Contributions to stability theory.
\newblock \emph{Annals of Mathematics}, 64\penalty0 (1):\penalty0 182--206,
  1956.

\bibitem[Mironchenko and Wirth(2018)]{mirwir_tac18}
A.~Mironchenko and F.~Wirth.
\newblock Characterizations of input-to-state stability for
  infinite-dimensional systems.
\newblock \emph{IEEE Trans. on Automatic Control}, 62\penalty0 (6):\penalty0
  1692--1707, 2018.

\end{thebibliography}
%\bibliography{C:/Users/Jose/Dropbox/biblio_latex/strings.bib,C:/Users/Jose/Dropbox/biblio_latex/complete_v2.bib,C:/Users/Jose/Dropbox/biblio_latex/hernan_v2.bib}
\end{document}